\documentclass[twocolumn,pra,showpacs,superscriptaddress]{revtex4}

\usepackage{graphicx}
\usepackage{dcolumn}
\usepackage{bm}

\begin{document}

\title{Particle number fractionalization of a one-dimensional atomic Fermi gas with synthetic spin-orbit coupling}

\author{Dan-Wei Zhang}
\affiliation{Laboratory of Quantum Information Technology and SPTE,
South China Normal University, Guangzhou 510006, China}
\affiliation{Department of Physics and Center of Theoretical and
Computational Physics, The University of Hong Kong, Pokfulam Road,
Hong Kong, China}

\author{L.-B. Shao}
\affiliation{Department of Physics and
Center of Theoretical and Computational Physics, The University of
Hong Kong, Pokfulam Road, Hong Kong, China}

\author{Zheng-Yuan Xue}
\affiliation{Laboratory of Quantum Information Technology and SPTE,
South China Normal University, Guangzhou 510006, China}

\author{Hui Yan}
\affiliation{Laboratory of Quantum Information Technology and SPTE,
South China Normal University, Guangzhou 510006, China}

\author{Z. D. Wang}
\email{zwang@hku.hk}\affiliation{Department of Physics and Center of
Theoretical and Computational Physics, The University of Hong Kong,
Pokfulam Road, Hong Kong, China}

\author{Shi-Liang Zhu}
\email{slzhu@scnu.edu.cn} \affiliation{Laboratory of Quantum
Information Technology and SPTE, South China Normal University,
Guangzhou 510006, China}
\affiliation{Department of Physics and
Center of Theoretical and Computational Physics, The University of
Hong Kong, Pokfulam Road, Hong Kong, China}


\date{\today}

\begin{abstract}

We propose an experimental scheme to simulate the
fractionalization of particle number by using a one-dimensional
spin-orbit coupled ultracold fermionic gas. The wanted spin-orbit
coupling, a kink-like potential, and a
conjugation-symmetry-breaking mass term are properly constructed
by laser-atom interactions, leading to an effective low-energy
relativistic Dirac Hamiltonian with a topologically nontrivial
background field. The designed system supports a localized soliton
excitation with a fractional particle number that is generally
irrational and experimentally tunable, providing a direct
realization of the celebrated generalized-Su-Schrieffer-Heeger
model. In addition, we elaborate on how to detect the induced
soliton mode with the fractional particle number in the system.

\end{abstract}


\maketitle

\section{introduction}

The idea of fractional particle number (FPN) goes back to the
Jackiw-Rebbi model \cite{Jackiw-Rebbi,Niemi} in the relativistic
quantum field theory, where fractionalization of fermion number
exhibits when a fermionic field is coupled to a topologically
nontrivial background field. The first physical demonstration of
this remarkable phenomenon was proposed by Su, Schrieffer, and
Heeger (SSH), in which a domain wall in one-dimensional (1D)
dimerized polymers, such as polyacetylene, induces a zero-energy
soliton state \cite{SSH}. The particle-hole ambiguity of the zero
mode restricts the fractional fermion number to be only
$\pm\frac{1}{2}$ in this system \cite{hajime,campbell}. Afterwards,
achievements have been made to generalize it to an irrational
fermion number by introducing another field to break the conjugation
symmetry, such as different on-site energies
\cite{Goldstone,Jackiw1983,rice}.

Another famous example of FPN is illustrated in the fractional
quantum Hall effect regime \cite{Laughin}, where the Laughin
quasiparticles not only have fractional charges but also have
fractional (anyonic) statistics in two dimensions (2D). Recent
search for fractionlization in 2D systems has theoretically
demonstrated that fractionally charged excitations may exist in
graphenelike \cite{Chamon}, square-lattice \cite{Seradjeh} and
kagome-lattice \cite{Ruegg} systems with vortex-type order
parameters (which describe the mass of the analog Dirac fermions in
the systems). The newly discovered quantum spin Hall insulators were
also proposed for realizing the SSH model based on the proximity
effect, which introduces a magnetic domain-wall
\cite{Qi2008,Vayrynen}. Notably, the edge electrons there with the
inherent chiral symmetry may exhibit a direct signature of FPN
\cite{Qi2008}.

On the other hand, quantum simulation of relativistic Dirac
Hamiltonian by using ultracold atomic gases has recently attracted
great interest \cite{Zhang}. For example, ultracold fermionic atoms
trapped in a honeycomb optical lattice (OL) were theoretically
proposed to behave as massless and massive Dirac fermions
\cite{Zhu2007}, and confirmed in a recent experiment
\cite{Tarruell}. The atomic gases with the synthetic spin-orbit (SO)
coupling \cite{Lin,Chen2012,Wang2012,Cheuk2012} through
light-induced gauge fields \cite{Dalibard,Wilczek} were also
proposed for investigating interesting Dirac dynamics
\cite{Vaishnav,Merkl,cwzhang,zhang2,Zhu2009,Goldman}. These cold
atom systems provide a highly controllable platform for studying a
wide range of models in relativistic quantum mechanics and field
theory \cite{Zhang}. Interestingly, Ruostekoski {\sl et al.}
presented an experimental scheme to realize \cite{Ruostekoski2002}
and detect \cite{Ruostekoski2003} the fractionalization of particle
number by using a two-component ultracold Fermi gas in a 1D optical
superlattice. The low-energy effective theory for the atoms in the
system becomes relativistic under certain conditions, and a
laser-induced modulation of atomic hopping between neighbor lattices
with a kink profile gives rise to a physical domain-wall, leading to
soliton modes with FPN.

Inspired by recent experimental achievements in the artificial SO
coupling in ultracold bulk bosonic \cite{Lin,Chen2012} and
particularly fermionic atoms \cite{Wang2012,Cheuk2012}, we here
present an alternative proposal for realizing the particle number
fractionalization using a 1D atomic Fermi gas with the synthetic SO
coupling. The required SO interactions and a kink-like potential are
properly constructed by dressing atoms with laser beams in the
system, such that the low-energy fermionic atoms can behave as
massless Dirac fermions coupling to a topologically nontrivial
background field. As a result, a localized soliton excitation in the
middle of the effective energy gap appears on the domain wall, which
is a direct quantum simulation of the standard SSH model. Another
two laser beams are used to introduce an effective Zeeman term,
which shifts the soliton excitation from the zero-energy. For a
midgap state below the zero energy level, it takes more fractional
fermion number from the valance band and less from the conduction
band, and vice versa for the opposite case, such that the soliton
state exhibits an irrational FPN in this general case; moreover, its
profile and FPN in the system are experimentally tunable.
Furthermore, we suggest experimentally available methods to detect
the induced soliton modes with FPN through measuring the soliton
density distribution and the local density of states (LDOS) near the
kink. We also discuss the possibility of generalizing our proposal
to the realization of the FPN in higher spatial dimensions.

The paper is organized as follows. In the next section (Sec. II), we
propose an experimental scheme to simulate the generalized SSH model
with an irrational FPN by using a 1D SO-coupled atomic Fermi gas.
The realization of a relativistic Dirac Hamiltonian with a kink
background field is shown, and the induced FPN in the system is
calculated and explained. In Sec. III, we elaborate on how to detect
the soliton modes with FPN in experiments. Finally, in Sec. IV, we
briefly discuss the generalization of the system to higher
dimensions and present conclusions.

\section{simulation of FPN with SO coupled fermionic atoms}

In this section, we show how to simulate the fractionalization of
particle number by using atomic Fermi gases with the synthetic
spin-orbit coupling. Let us start with a brief review of the
celebrated model describing kink-soliton states and arbitrary
fractional fermion number in the context of relativistic quantum
field theory \cite{Jackiw-Rebbi,Goldstone}. For 1D massless Dirac
fermions subject to two static bosonic fields $\varphi_1$ and
$\varphi_2$, the relativistic Dirac Hamiltonian is given by
\cite{note1}
%
%
%
\begin{equation}
\label{Dirac_H}
 H_D=c\sigma_x p_x-\varphi_2(x)\sigma_y+\varphi_1(x)\sigma_z,
\end{equation}
where $c$ is the effective speed of light and $\sigma_{x,y,z}$ are
the Pauli matrices. The background field with a kink potential is
described by \cite{Goldstone,Jackiw1983}
\begin{equation}
\label{kink} \varphi_1(x)=\varphi_1^0,~~
\varphi_2(x\rightarrow\pm\infty)=\pm\varphi_2^0,
\end{equation}
where $\varphi_1^0$ and $\varphi_2^0$ are constants. The kink
$\varphi_2$  acts as the boundary of two degenerate vacuums
\cite{Jackiw1983}. The relativistic Dirac Hamiltonian with such a
topologically nontrivial background potential supports an unpaired
soliton state, which gives rise to fractionalization of particle
number \cite{Goldstone}. Moreover, the FPN is generally irrational
and takes one-half in the standard SSH model with the
conjugation-symmetry when $\varphi^0_1$ is vanishing.

\begin{figure}[tbph]
\vspace{0.5cm}
\label{Fig1} 
\includegraphics[height=3cm,width=8cm]{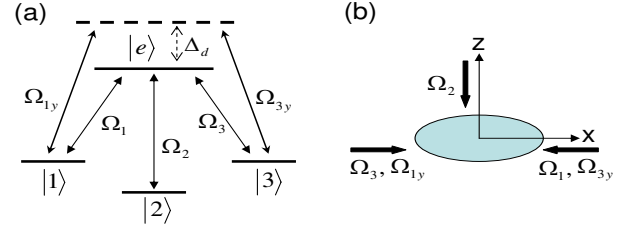}
\caption{(Color online) Schematic representation of the laser-atom
interaction configuration for generating an effective relativistic
Dirac Hamiltonian. (a) The three ground states $|1\rangle$,
$|2\rangle$ and $|3\rangle$ are resonantly coupled to the excite
states $|e\rangle$ by lasers with Rabi frequencies $\Omega_j$, and
two additional lasers $\Omega_{1y}$ and $\Omega_{3y}$ couples
$|1\rangle$ and $|3\rangle$ to $|e\rangle$ with a large detuning
$\Delta_d$. The choose for the three ground states is that,
$|1\rangle$ and $|3\rangle$ are two degenerate Zeeman sublevels
which are addressed by laser beams with different polarization and
$|2\rangle$ is another hyperfine level with different energy so that
it is addressed by a laser with a different frequency. (b) The
spatial configuration and propagating direction of the laser beams.
The candidate for the fermionic atoms can be $^6$Li or $^{40}$K.}
\end{figure}

\subsection{Realizing the relativistic Dirac Hamiltonian in cold atom systems}

Now we demonstrate how to realize the wanted Dirac Hamiltonian
(\ref{Dirac_H}) with a SO-coupled atomic Fermi gas. We consider an
ensemble of quasi-2D noninteracting fermionic atoms with three
relevant spin components in the ground-state manifold
$\{|1\rangle, |2\rangle, |3\rangle\}$, which are resonantly
coupled to a common excited state $|e\rangle$ through the standard
tripod configuration \cite{Ruseckas,zhu2011} as shown in Fig. 1. A
candidate for the fermionic atoms can be $^6$Li or $^{40}$K. For
$^6$Li atoms, the hyperfine levels can be selected as
\begin{equation}
\label{Li6}
\begin{array}{ll}
|1\rangle=|2^2{\text
S}_{1/2},F=\frac{3}{2},m_F=-\frac{1}{2}\rangle,\\\\
|2\rangle=|2^2{\text
S}_{1/2},F=\frac{1}{2},m_F=\frac{1}{2}\rangle,\\\\
|3\rangle=|2^2{\text
S}_{1/2},F=\frac{3}{2},m_F=\frac{3}{2}\rangle,\\\\
|e\rangle=|2^2{\text P}_{1/2},F=\frac{1}{2},m_F=\frac{1}{2}\rangle.
\end{array}
\end{equation}
For $^{40}$K atoms, the corresponding hyperfine levels can be
\begin{equation}
\label{K40}
\begin{array}{ll}
|1\rangle=|4^2{\text
S}_{1/2},F=\frac{7}{2},m_F=-\frac{1}{2}\rangle,\\\\
|2\rangle=|4^2{\text
S}_{1/2},F=\frac{9}{2},m_F=\frac{1}{2}\rangle,\\\\
|3\rangle=|4^2{\text
S}_{1/2},F=\frac{7}{2},m_F=\frac{3}{2}\rangle,\\\\
|e\rangle=|4^2{\text
P}_{1/2},F=\frac{9}{2},m_F=\frac{1}{2}\rangle.\\\\
\end{array}
\end{equation}
The corresponding Rabi frequencies of the three resonantly
coupling laser beams can be parameterized as
\begin{eqnarray}
\nonumber \Omega _{1}&=&\Omega \,\sin \alpha \,\cos \theta
\,\mathrm{e}^{-i\kappa x},\\ \label{Rabi}
\Omega _{2}&=&\Omega \,\cos \alpha \,%
\mathrm{e}^{-i\eta\kappa z},\\ \nonumber \Omega _{3}&=&\Omega \,\sin \alpha \,\sin \theta \,%
\mathrm{e}^{i\kappa x}.\end{eqnarray}
 The wave numbers are
$\kappa$ and $\eta\kappa$ as shown in Fig. 1(a), and $\Omega
=\sqrt{\sum_{j=1}^{3}|\Omega _{j}|^{2}}$ is the total Rabi
frequency. Here $\eta=1+\delta\eta$ with deviation $\delta\eta$
being for matching the resonant-coupling frequency of the second
laser beam. For the selected atomic hyperfine states in Eqs.
(\ref{Li6}) and (\ref{K40}), $\delta\eta\approx5\times10^{-7}$ for
$^6$Li atoms and $\delta\eta\approx3.5\times10^{-6}$ for $^{40}$K
atoms. This can be  achieved in experiments by adjusting the laser
frequency. The deviation is negligible in our derivations, however,
we still use the notation $\eta$ in the following discussions for
consistency. We further adopt uniform plane-wave laser beams that
$\Omega$, $\alpha$, and $\theta$ are all constants, and particularly
choose $\theta=\pi/4$.

\begin{figure}[tbph]
\vspace{0.5cm}
\label{Fig2} 
\includegraphics[height=4cm,width=8.5cm]{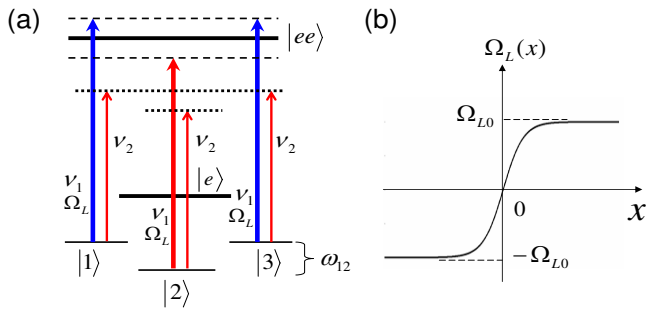}
\caption{(Color online) Schematic representation of (a) the coupling
lasers which generate the wanted external potentials $V_{1,2,3}$ in
Eq. (\ref{potential}); and (b) the spatial configuration of the Rabi
frequency $\Omega_L(x)$ which forms the needed kink-like potential.
The frequency difference $\omega_{12}=228$ MHz for $^6$Li atoms and
$\omega_{12}=1186$ MHz for $^{40}$K atoms for the selected hyperfine
levels in the text, both of which are much larger than the natural
linewidth of the excited state $|ee\rangle$ (about 6 MHz). }
\end{figure}

The single-particle Hamiltonian for each atom takes the form $H_{s}=\mathbf{p%
}^{2}/2m_a+H_{\text{int}}$, where $\mathbf{p}$ denotes the momentum
operator and $m_a$ is the atomic mass. The light-atom interaction
Hamiltonian $H_{\text{int}}$ is given by $H_{\text{int}}=\hbar
\sum_{j=1}^{3}(\Omega _{j}|e\rangle \langle j|+{\text{H.c.}})$.
Diagonalizing $H_{\text{int}}$
yields two orthogonal dark states
\begin{equation}
\begin{array}{ll}
\label{spinstates}
|D_{1}\rangle = \left(\mathrm{e}%
^{i\kappa x}|1\rangle -
\mathrm{e}^{-i\kappa x}|3\rangle\right)\mathrm{e}^{-i\eta\kappa z}/\sqrt{2}, \\
|D_{2}\rangle = \left(\mathrm{e}%
^{i\kappa x}|1\rangle + \mathrm{e}^{-i\kappa
x}|3\rangle\right)\mathrm{e}^{-i\eta\kappa z}\cos
\alpha/\sqrt{2}-\sin \alpha |2\rangle.
\end{array}
\end{equation}
%
The dark states are decoupled to the excited state $|e\rangle$ and
thus are immune to spontaneous emission. They span a degenerate
subspace, in which the full state of a single atom can be written as
$|\chi(\textbf{r})\rangle =
\sum_{i=1}^{2}\psi_{i}(\textbf{r})|D_i(\textbf{r})\rangle$. The
center-of-mass amplitudes $\psi_i(\textbf{r})$ corresponding to the
spatial wave functions of the two dark states obey the
Schr\"{o}dinger equation $i\hbar
\partial_t \Phi = H \Phi$, where the two-component spinor $\Phi({\bf r})=(\psi_1({\bf r}),\psi_2({\bf r}))^T$
and the Hamiltonian reads
\begin{equation}
\label{FullHamiltonian} H = \frac{1}{2m_a}(\textbf{p} -
\textbf{A})^2 + \phi + V .
\end{equation}
The gauge potential $\bf{A}$ arises from the position dependence of
the dark states and is given by $\textbf{A}_{j,n} = i\hbar\langle
D_j(\textbf{r})|\nabla|D_n(\textbf{r})\rangle$
\cite{Dalibard,Wilczek}. The projecting-induced scalar potential
$\phi$ and the external potential $V$ are respectively determined by
$\phi_{j,n}=\sum_{l=1}^2\vec{A}_{j,l}\vec{A}_{l,n} /2m_a$ and
$V_{j,n} = \langle D_j(\textbf{r})|\hat{V}|D_n(\textbf{r})\rangle$
with $\hat{V}=\sum_{l=1}^{3}V_{l}(\textbf{r})|i\rangle\langle i|$
\cite{Dalibard,Wilczek}. We consider the adiabatic motion of atoms
initially prepared in the dark state subspace. It should be noted
that the two dark states are not the lowest energy states in this
system so that the adiabatic approximation works well only for
finite time scales (mainly due to collisional relaxations), up to
several hundred milliseconds under realistic conditions
\cite{zhu2011}. To obtain a lowest-energy two-fold (nearly)
degenerate subspace, one can adopt the optical dressing scheme
described in Ref. \cite{zhang2} (where atoms with simpler $\Lambda$
type configuration) or in Ref. \cite{Juzeliunas} (where more atomic
internal states and coupling lasers are required).

The Rabi frequencies chosen in Eq.(\ref{Rabi}) can realize the first
term in Hamiltonian (\ref{Dirac_H}) with certain potentials;
however, to implement the required potentials exactly in
Eq.(\ref{Dirac_H}), more complicated laser configures are needed.
One possible method to generate the required potentials is that we
further choose two additional laser beams with frequencies $\nu_1$
and $\nu_2$, as schematically shown in Fig. 2(a). The first laser
beam (denoted by frequency $\nu_1$) with  the effective Rabi
frequency $\Omega_L$ (which takes the real Rabi frequency and the
detuning into account) is blue detuned for atoms in the internal
levels $|1\rangle$ and $|3\rangle$, but red detuned for atoms in
level $|2\rangle$, all of which are far-off-resonantly coupled to
another excited state $|ee\rangle$. This energy state can be
selected as $|ee\rangle=|2^2{\text
P}_{3/2},F=\frac{3}{2},m_F=\frac{1}{2}\rangle$ for $^6$Li atoms and
$|ee\rangle=|4^2{\text
P}_{3/2},F=\frac{9}{2},m_F=\frac{1}{2}\rangle$ for $^{40}$K atoms,
respectively. In addition, the second laser (denoted by frequency
$\nu_2$), which is also far-off-resonant, is use to create constant
energy terms in $V_{1,3}$ and $V_2$ in the following equation
(\ref{potential}). The energy difference between them [cf. Eq.
(\ref{potential})] can be realized by detuning the second laser from
the two-photon resonance with the frequency
$\hbar\kappa^2(1-\eta^2\cos^2\alpha)/2m_a$. Thus the resulting
external potentials are given by
\begin{equation}
\label{potential}
\begin{array}{ll}
V_1=V_3=\hbar\Omega_L(x,z)-\frac{\hbar^2\kappa^2}{2m_a},\\ \\
V_2=-\hbar\Omega_L(x,z)-\frac{\hbar^2\eta^2\kappa^2}{2m_a}\cos^2\alpha.
\end{array}
\end{equation}

After introducing all of the laser configurations, we can obtain the
total resulting potentials for the atoms in the laser field as
\begin{equation}
\begin{array}{ll}
\textbf{A}  = -\hbar \kappa\cos\alpha\ \sigma_x \vec{e}_x + \hbar
\eta\kappa \left(\begin{array}{cc}1 & 0 \\0 & \cos^2\alpha
\end{array}\right) \vec{e}_z,\\ \\ \phi  = \frac{\hbar^2\kappa^2\sin^2\alpha
}{2m_a}\left(\begin{array}{cc} 1 & 0 \\0 &
\eta^2\cos^2\alpha\end{array}\right),\\ \\ V  =
\left(\begin{array}{cc}V_1 & 0 \\0 & V_1\cos^2\alpha +
V_2\sin^2\alpha\end{array}\right),
\end{array}
\end{equation}
with $\phi+V=\hbar\Omega_L\sin^2\alpha\sigma_z$ up to an
irrelevant constant. Note that atoms in such a synthetic
non-Abelian gauge field behave as electrons with a SO coupling,
which can be seen from the term $\textbf{p}\cdot\textbf{A}$ in
Hamiltonian (\ref{FullHamiltonian}).

By applying an additional extremely anisotropic trapping potential
to freeze the atomic motions along $z$ axis, we arrive at the
quasi-1D cases \cite{note2}. For ultralow temperature, the
momentum of atoms along $x$ axis $p_x \ll\hbar\kappa\cos\alpha$,
such that the $p_x^2$ term in Eq. (\ref{FullHamiltonian}) may be
safely neglected, leading to an effective Dirac Hamiltonian
\begin{equation}
\label{JackiwHam} H_{e} \approx c_x \sigma_x p_x+ \Delta(x)
\sigma_z,
\end{equation}
where $c_x=\hbar\kappa\cos\alpha/m_a$ is the effective speed of
light in this system and $\Delta(x)=\hbar\Omega_L(x)\sin^2\alpha$.
The Hamiltonian (\ref{JackiwHam}) describes a massive Dirac
fermion having a position-dependent mass $\Delta(x)/c_x^2$, or in
another point of view, a massless Dirac fermion coupling to a
static background field $\Delta(x)$ \cite{Jackiw-Rebbi}. If we
choose the intensity distribution of the laser beam with photon
frequency $\nu_1$  as a kink-type function along $x$ axis, then
the standard SSH model in continuum limit \cite{SSH} is realized
in this cold atom system. It is interesting to note that the
laser-atom interaction of $\Lambda$ configuration \cite{zhang2}
can also be used to realize the relativistic Dirac Hamiltonian
(\ref{JackiwHam}), and in this case the experimental setup can
even be simpler. While it is noted that such a simplified scheme
is unable to be extended to realize the generalized SSH model
described in Eq. (\ref{ModelHam1}) below with an irrational FPN.

To introduce the constant field $\varphi^0_1$ in Hamiltonian
(\ref{Dirac_H}), which acts as a mass term and breaks the
conjugation symmetry, we can apply two additional laser beams to
couple the atomic states $|1\rangle$ and $|3\rangle$ to the excited
state $|e\rangle$ off-resonantly with a large detuning $\Delta_d$ as
shown in Fig. 1, with the corresponding Rabi frequencies $\Omega
_{1y}=i\left\vert \Omega _{y}\right\vert \mathrm{e}^{i\kappa x}$ and
$\Omega _{3y}=\left\vert \Omega _{y}\right\vert \mathrm{e}^{-i\kappa
x}$ \cite{zhu2011}. Since $\left\vert \Omega _{1y}\right\vert
,\left\vert \Omega _{3y}\right\vert \ll \Delta_d $, the effective
Hamiltonian describes a perturbation coupling between states
$|1\rangle$ and $|3\rangle$, which is given by $H_{p}=-i\hbar \Omega
_{p}e^{2i\kappa x}|1\rangle \langle 3|+{\text{H.c.}}$ with $\Omega
_{p}=|\Omega _{y}|^{2}/\Delta_d $ \cite{note3}. We assume $\Omega
_{p}\ll \Omega $, so the Hamiltonian $H_{p}$ can not pump the atoms
outside of the dark state subspace. Mapping $H_{p}$ into the
subspace spanned by the basis $\{|D_1 \rangle, |D_2 \rangle \}$, we
obtain $H_{p}= \hbar \Omega _{p}\cos \alpha \sigma _{y}$. Therefore,
the total 1D effective Hamiltonian for the ultracold atoms is
${H}_{\text{1D}} = H_e+H_p$. By introducing a unitary transformation
$\Phi(x)\rightarrow e^{i\frac{\pi}{4}\sigma_{x}}\Phi(x)$, we can
obtain the Dirac Hamiltonian
\begin{equation}
\label{ModelHam1} {H}_{\text{1D}} =  c_x \sigma_x p_x + \Delta(x)
\sigma_y-\Gamma\sigma_z,
\end{equation}
where $\Gamma = \hbar|\Omega_y|^2\cos\alpha/\Delta_d$. We note that
the unitary transformation used here is for mathematical convenience
but involves no manipulation on the system. Compared to the original
Dirac Hamiltonian (\ref{Dirac_H}), here the effective field
$-\Gamma$ corresponds to the constant background $\varphi^0_1$, and
the field $\Delta(x)$ should present a kink-like profile, which
corresponds to $-\varphi_2(x)$. To this end, we can choose the
spatial profile of the Rabi frequency $\Omega_L(x)$ with the kink
form as shown in Fig. 2(b), and $\Omega_L(x)=\pm \Omega_{L0}$ as
$x\rightarrow\pm\infty$. Thus the asymptotic value of $\Delta(x)$
can be denoted as
$\Delta_0\equiv\Delta(x\rightarrow+\infty)=\hbar\Omega_{L0}\sin^2\alpha$.

Finally in this section, we note that the recent experiment of
generating SO couplings in Fermi gases \cite{Wang2012} may also be
extended to realize the Dirac Hamiltonians (\ref{JackiwHam}) and
(\ref{ModelHam1}). In the experiment \cite{Wang2012}, two spin-1/2
states are chosen as two internal hyperfine states instead of
dressed states [see Eq. (\ref{spinstates})] in our scheme, and they
are coupled by a pair of Raman beams with spatially homogenous
coupling strength $\Omega_{R}$. The synthetic SO coupling is just
one-dimensional with the form $p_x\sigma_z$, and there is an
additional term related to the Raman coupling $\Omega_{R}\sigma_x$
in the single-particle Hamiltonian (see Refs. \cite{Lin,Wang2012}
for details). If one uses Raman beams with spatially imhomogenous
coupling strength and kink-type profile along $x$ axis [i.e.
$\Omega_{R}(x)=\Omega_{L}(x)$], the low-energy effective Hamiltonian
for the atoms takes the form of the Dirac Hamiltonian
(\ref{JackiwHam}) under a spin rotation. To simulate the generated
SSH model Hamiltonian (\ref{ModelHam1}), one needs additional laser
beams or radio-frequency fields to couple the two spin states as a
$\sigma_y$-coupling term.

\subsection{Fractional particle number in this system}

We now turn to calculate the FPN in the proposed system. There is a
number of methods for computing the FPN of topological solitons
\cite{Niemi}, including the well-known conjugation-symmetry analysis
for the zero modes with one-half fermion number
\cite{Niemi,Jackiw2}. It was first pointed out by Goldstone and
Wilczek that \cite{Goldstone}, at zero temperature, the fractional
fermion number of the soliton in this model is determined by the
kink background field (\ref{kink}). The adiabatic condition was
imposed there for a valid perturbation calculation by assuming
$|\partial\varphi_{i}|\ll m^2$ ($i=1,2$), where
$m\equiv\sqrt{\varphi_{1}^{2}+\varphi_{2}^{2}}$. However, Yamagishi
showed that the exact result actually does not depend on the
adiabatic condition \cite{Yamagishi}. For simplicity, we here employ
still the adiabatic condition to derive the result in a new but
simple way. The current of this (1+1)D system is
\begin{equation}
j^{\mu}(x)=-\langle
x|\mathbf{tr}\gamma^{\mu}\widehat{G}|x^{+}\rangle,
\end{equation}
where the Dirac matrices $\gamma^0=\sigma_z$,
$\gamma^1=i\sigma_y$, $\gamma^5=\sigma_x$, and
the Green's function of the relativistic Dirac Hamiltonian
(\ref{Dirac_H}) is given by
\begin{eqnarray}
\label{Gfunc1}
G=\frac{i}{\gamma^{\mu}\hat{p}_{\mu}-(\varphi_{1}+i\varphi_{2}\gamma^{5})}.
\end{eqnarray}
Here $\mu=0,1$ correspond to the time and space components,
respectively.

In the derivation, we have used the adiabatic approximation that
$\partial\varphi_{i}$ commutes with $\frac{1}{\hat{p}^{2}-m^{2}}$
and kept the first order approximation. Thus the Green's function
can be written as
%
\begin{equation}
\begin{array}{ll}
\label{Gfunc2}
G=[\gamma^{\mu}\hat{p}_{\mu}+(\varphi_{1}-i\varphi_{2}\gamma^{5})]\{G^{-1}[\gamma^{\mu}\hat{p}_{\mu}+(\varphi_{1}-i\varphi_{2}\gamma^{5})]\}^{-1}\\\\
~~\approx\left[\gamma^{\mu}\hat{p}_{\mu}+(\varphi_{1}-i\varphi_{2}\gamma^{5})\right]\frac{i}{\hat{p}^{2}-m^{2}}\\\\
~~~~~-\left[\gamma^{\mu}\hat{p}_{\mu}+(\varphi_{1}-i\varphi_{2}\gamma^{5})\right]
\frac{1}{(\hat{p}^{2}-m^{2})^{2}}\gamma^{\nu}\partial_{\nu}(\varphi_{1}-i\varphi_{2}\gamma^{5}).
\end{array}
\end{equation}
After a straightforward calculation, given that the chemical
potential (the Fermi level) is zero, we can obtain the average
current in the background field as \cite{Goldstone}
\begin{equation}
j^{\mu}(x)=-\frac{1}{2\pi}\epsilon^{\mu\nu}\partial_{\nu}\Theta(x),
\end{equation}
where $\Theta(x)=\arg(\varphi_{1}+i\varphi_{2})$ denotes the angular
field of the background, and $\epsilon^{\mu\nu}$ is the two-index
totally antisymmetric tensor. When the chemical potential
$\tilde{\mu}=0$, the particle density is given by
$\rho_0(x)=-\frac{1}{2\pi}\frac{\partial\Theta(x)}{\partial x}$ with
respect to the density of kink-free system.
Thus we obtain the FPN $\mathcal{N}_0=\int \rho_0(x) dx$ in this
system with $\varphi_{1}^0=-\Gamma$ and $\varphi_{2}(x)=-\Delta(x)$
as
\begin{equation}
\label{FPN}
\mathcal{N}_0=\frac{1}{\pi}\arctan\left(\frac{\Delta_0}{-\Gamma}\right)=-\frac{1}{\pi}\arctan\left(\frac{\Omega_{L0}\Delta_d\sin^2\alpha}{|\Omega_y|^2\cos\alpha}\right).
\end{equation}

It is clear from Eq. (\ref{FPN}) that $\mathcal{N}_0$ is generally
irrational and can be an arbitrary fractional number in the range
$(-\frac{1}{2},\frac{1}{2})$ for a finite $\Gamma$. Especially, the
conjugation-symmetric Jackiw-Rebbi model is obtained in the limit
$\Gamma\rightarrow0$, i.e., without applying the laser beams
$\Omega_{{\text 1y}}$ and $\Omega_{{\text 3y}}$. In this case, the
soliton is a zero-energy mode with one-half fermion number
$\mathcal{N}_0=\pm\frac{1}{2}$. It is interesting to note that the
FPN in this system is widely tunable in experiments via laser-atom
interactions [cf. Eq. (\ref{FPN})], making it a controllable
platform for simulating fractionalization of particle number.

\begin{figure}[tbph]
\vspace{0.5cm}
\label{Fig2} 
\includegraphics[height=6.5cm,width=8.5cm]{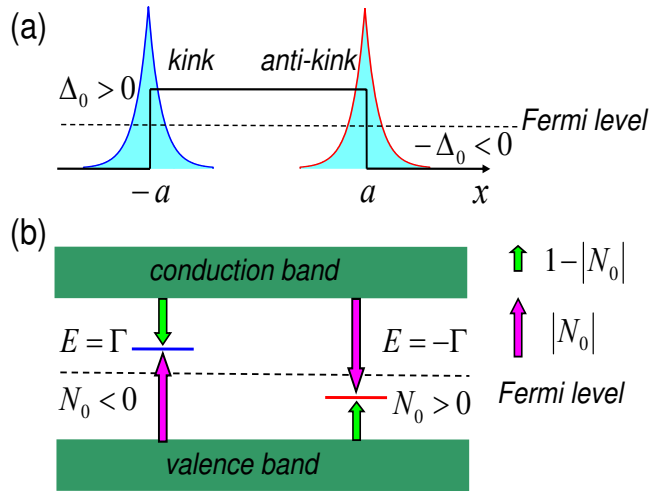}
\caption{(Color online) Schematic representation of (a) a background
field with a pair of kink and anti-kink, both of which support a
localized soliton state with profile $\sim\exp{(-\Delta_0|x\pm
a|/\hbar c_x)}$; and (b) the energy spectrum and a pair of solitons
with energies $E=\pm \Gamma$ and fractional particle numbers
$\pm|\mathcal{N}_0|$. The midgap soliton modes are protected by an
energy gap $E_g=2\sqrt{\Delta^2_0+\Gamma^2}$ between the effective
conduction band and valence band. At the kink, the soliton state
picks a FPN of $|\mathcal{N}_0|$ from the effective valance band and
another $1-|\mathcal{N}_0|$ from the effective conduction band, and
vice versa for the opposite case at the anti-kink.}
\end{figure}

Fractionalization has been widely investigated in relativistic
quantum field theory \cite{Jackiw,Niemi} and condensed matter
systems
\cite{SSH,Goldstone,Jackiw1983,rice,Laughin,Chamon,Seradjeh,Ruegg,Qi2008,Vayrynen},
where it can give rise to interesting transport phenomena. For
example, the existence of fractionally charged excitations greatly
enhances the conductivity in the polymers \cite{SSH} and may induce
quantized currents in the quantum spin Hall insulators
\cite{Qi2008}. The interpretation of FPN in these kink-soliton
systems is usually in terms of deformation (or polarization) of the
ground-state vacuum due to the kink which supports a single soliton
mode \cite{Niemi,Jackiw}. This fractionalization mechanism is very
different from that in the fractional quantum Hall effect regime,
where the fractional collective excitations are described by a
strongly correlated Laughlin wave function \cite{Laughin}.

To have better understanding of this mechanism, we consider another
kind of background field with a simple but experimentally practical
configuration, that is, a pair of kink and anti-kink both with a
step-function profile as shown in Fig. 3(a). Here we assume
$\Delta_0>0$ and $a\gg\hbar c_x/\Delta_0$ such that the kink and
anti-kink are almost decoupled. By solving the energy spectrum of
Hamiltonian (\ref{ModelHam1}) at the kink potential (near $x=-a$)
with $\Delta(x)=\Delta_{0}\textbf{sgn}(x+a)$, we find that there is
a localized midgap eigenstate in the kink at $E=\Gamma$ with the
wavefunction decaying as $\exp{(-\Delta_0|x+a|/\hbar c_x)}$ and the
energy gap $E_g=2\sqrt{\Delta^2_0+\Gamma^2}$ . It is understood that
the isolated state picks up a fractional fermion number (i.e., FPN)
of $|\mathcal{N}_0|$ [see Eq. (\ref{FPN})] from the effective
valance band and $(1-|\mathcal{N}_0|)$ from the effective conduction
band, as shown in Fig. 3(b). For an anti-kink potential (near $x=a$)
with $\Delta(x)=-\Delta_{0}\textbf{sgn}(x-a)$, the localized soliton
state is obtained at $E=-\Gamma$ with the wavefunction decaying as
$\exp{(-\Delta_0|x-a|/\hbar c_x)}$ as shown in Fig. 3(a). It picks
up $(1-|\mathcal{N}_0)|$ from the valence band and $|\mathcal{N}_0|$
from the conduction band. For a periodic system, there must be pairs
of kink and anti-kink. If both states are unoccupied, the particle
numbers are $-|\mathcal{N}_0|$ at the kink and $|\mathcal{N}_0|-1$
at the anti-kink. When the chemical potential, i.e., the effective
Fermi level in this system, is tuned up, the $E=-\Gamma$ soliton
state is occupied first and the particle numbers at kink and
anti-kink are $\mp|\mathcal{N}_0|$, respectively. And when both
states are occupied, there are particles $(1-|\mathcal{N}_0|)$ and
$|\mathcal{N}_0|$ at the kink and anti-kink.

From Eq. (\ref{FPN}), we can see that the FPN $\mathcal{N}_0$
depends only on the asymptotic value of the kink $\Delta_0$
rather than the detailed shape of $\Delta(x)$. In this sense, it is
topological and is insensitive to local fluctuations of the
background field. This property enables us to use laser beams of
different and imperfect intensity distributions compared to the
ones with the exact kink profile as shown in Fig. 2(b) and with a wide
square-potential profile of a kink-anti-kink pair as shown in Fig.
3(a). In experiments, the intensity distribution of laser beams
can be well designed and the wanted ones with nearly
square-potential profiles have been realized \cite{Tarallo}.
Although the value of $\mathcal{N}_0$ is obtained at zero
temperature, the corresponding FPN for finite temperature $T$
defined as $\mathcal{N}_T$ can also be calculated by taking the
thermal distribution (i.e., Dirac-Fermi distribution) into account
\cite{Dunne}. Interestingly, $\mathcal{N}_T$ is just depends on the
asymptotic value of the background field and the temperature $T$
\cite{Dunne}. At low temperatures, i.e., $|\varphi^0_1|\beta\gg1$
with $\beta=1/k_BT$ ($k_B$ is the Boltzmann constant), one has
$\mathcal{N}_T\approx \mathcal{N}_0-e^{-|\varphi^0_1|\beta}
$\cite{Dunne}. For our proposed cold atom system with the typical
temperature $T\sim0.1$ $\mu$k and parameter $\Gamma/\hbar\sim0.1$
MHz, the deviation $\mathcal{N}_0-\mathcal{N}_T\approx
e^{-\Gamma\beta}\approx e^{-10}$. Thus we can conclude that the FPN
in this system is very robust against the finite-temperature
modification due to the gap protection.

Before ending this section, we discuss briefly the modifications on
the soliton state and its FPN arising from the neglected quadratic
term in the Dirac Hamiltonian (\ref{JackiwHam}). For this system of
bulk atomic gases, the effective cutoff momentum is determined by
the Fermi level and the typical atomic momentum can be one order
less than the recoil momentum of photons (for temperature $T\sim0.1$
$\mu$k and $\kappa\cos\alpha\sim10^7$ ${\text m}^{-1}$). So we can
treat it as a perturbation $\delta H=p_x^2/2m_a$. This perturbation
alters the energy spectrum and also breaks the conjugation symmetry,
but the Dirac point and the energy gap opened by the kink-background
remain. From the perturbation calculations, we find that both of the
spatial profile and the energy of the soliton state are modified.
For the case of step-function kink potential and when $\Gamma=0$,
the spatial wavefunction of the soliton state decays as
$\left[1+\frac{\Delta_0}{4m_ac^2_x}(1+\frac{\Delta_0|x+a|}{\hbar
c_x})\right]\exp{(-\Delta_0|x+a|/\hbar c_x)}$, which is slightly
broader than that in the absence of $\delta H$ [c.f. Fig. 3(a)]. The
corresponding energy is shifted from $E=0$ to
$E=-\Delta_0^2/2m_ac_x^2$ up to the first order perturbation. As
long as this energy shift is very small compared to the gap, i.e.
$\Delta_0/2m_ac^2_x\ll1$, one can expect that the soliton state is
robust against the breaking of conjugation symmetry induced by the
quadratic term \cite{Chamon}. In this case, the modification of FFN
can be estimated as
$\delta\mathcal{N}_0\approx\left|\frac{1}{2}-\frac{1}{\pi}\arctan(\frac{2m_ac^2_x}{\Delta_0})\right|$.
The situation is more complicated for $\Gamma\neq0$, but one can
still follow a similar perturbation argument and obtain the
corresponding modification of the soliton state at the kink.

\section{detection of soliton with FPN}

In this section, we propose possible methods for detecting the
fractionalization of particle number in the atomic system mainly
through the soliton density and the LDOS near the kink (or the
anti-kink) by using two standard experimental detection methods
for ultracold atomic gases, such as {\sl in-situ} absorption
imaging technique \cite{Bartenstein} and spatially resolved
radio-frequency (rf) spectroscopy \cite{Shin}.

First, the density distribution of the soliton modes may be
extracted out from the atomic density measurement via optical {\sl
in-situ} absorption imaging \cite{Bartenstein}. In this continuum
model, we work in the soliton framework, and the physical particle
number in the soliton sector is equivalently defined as being
measured relative to the free sector without the kink background
\cite{Jackiw1983,Jackiw}. Thus the density distribution of the
soliton mode is given by \cite{Jackiw1983,Jackiw}
\begin{equation}
\begin{array}{ll}
\rho_0(x) = \int dE
\left[|\Psi_E(x)|^2-|\psi_E(x)|^2\right]\\
\\~~~~~~~ =\left.\left[
\tilde{\Upsilon}(x)-\Upsilon(x)\right]\right|_{E_F=0},
\end{array}
\end{equation}
where $\Psi_E$ ($\psi_E$) and $\tilde{\Upsilon}(x)$ [$\Upsilon(x)$]
are the fermion single-particle energy eigenstates and atomic
density distribution in the presence (absence) of kink background
$\varphi_2(x)$, respectively. Note that here we have assumed the
effective Fermi level at $E_F=0$, which can be achieved by properly
tuning the chemical potential of the atomic gas. In this sense, we
can measure the spatial density distribution of the SO-coupled Fermi
gas both with and without the kink potential by tuning on and off
the laser beam $\Omega_L$, which correspond to $\tilde{\Upsilon}(x)$
and $\Upsilon(x)$, respectively. The integration of $\rho_0(x)$
gives the value of FPN $\mathcal{N}_0$ in Eq. (\ref{FPN}). This
detection scheme provides a clear physical picture of FPN; however,
it is hard to be implemented in a practical experiment, as there is
only one atom in the soliton sector (kink and anti-kink) comparing
to $N_a-1$ ones in the free sector, where the total number of
fermions $N_a$ is restricted by the chemical potential and is
usually several orders larger than unit. In addition, the number of
soliton modes can not be scaled with increasing $N_a$. However, the
occupation of the soliton state affects significantly the atomic
density distribution near the kink, which may be regarded as a
convenient feature to identify the existence of solitons.

An alternative but practical approach to probe the soliton state is
measuring the LDOS $\rho(x,E)$ near the kink (or anti-kink) by using
spatially resolved rf spectroscopy \cite{Shin}, which has been
proposed to detect other midgap bound states in bulk Fermi gases
\cite{Jiang}, including the zero-energy Majorana modes
\cite{zhu2011,Tewari,Liu}. The idea is that one first uses a probe
rf field to induce a single-particle excitation from the initial
state $|a_i\rangle$ to an unoccupied fluorescent probe state
$|a_f\rangle$, and then imaging the population in state
$|a_f\rangle$ to obtain the spatial information about the LDOS
\cite{Shin,Jiang,Liu}. If we assume that the probe field is weak and
is detuning $\delta_{\text{rf}}$ from the the induced transitions,
then the population change in state $|a_f\rangle$ can be calculated
from the linear response theory \cite{Jiang,Liu}
\begin{equation}
\label{STM} I(x,E)\equiv\frac{d}{dt}\langle a^{\dag}_f(x)
a_f(x)\rangle\propto\rho_{a_i}(x,E-\delta_{\text{rf}})\Xi(\delta_{\text{rf}}-E),
\end{equation}
where $\Xi(\cdot)$ is a unit step function. For a harmonically
trapped gas, the chemical potential changes from $\tilde{\mu}=0$
here to $\tilde{\mu}(x)=\frac{1}{2}m_a\omega^2x^2$ with $\omega$
being the trapping frequency under the local density approximation,
and $E$ in Eq. (\ref{STM}) is thus replaced by $E-\tilde{\mu}(x)$.
Due to the trapping potential, the energy variation over the length
scale $l_0\equiv\sqrt{E_g/m_a\omega^2}$ becomes comparable to the
energy gap which protects soliton modes. Therefore we could use a
sufficiently weak trap and experimentally control the size of the
gap to reach the nonvanishing gap and locally homogeneous limit.
Thus the previously presented physical picture about the soliton
mode persists. Since the soliton mode in the proposed system has
energy $\Gamma$ inside the gap and is localized at the kink $x=-a$,
there will be a significantly enhanced population transfer with
frequency $\delta_{\text{rf}}/\hbar=[\Gamma-\tilde{\mu}(-a)]/\hbar$
near the kink. The contribution from the soliton mode would be
clearly visible and well separated from other quasiparticle
contributions by the energy gap $E_g$. Thus the soliton density
distribution $\rho_0(x,\Gamma)$ can be mapped and singled out in
this way. Compared with the previous detection method, the later
scheme is insensitive to the fluctuations in the initial number of
fermions $N_a$ since (i) the occupation of the soliton mode just
depends on the Fermi level (i.e. the chemical potential at the kink)
and the fluctuations of $N_a$ will not affect the topology of the
Fermi level; (ii) the soliton mode is an eigenstate that is robust
against thermal and local fluctuations in the presence of an energy
gap.
Interestingly, even single atom in $|a_f\rangle$ state can be
detected with the standard quantum jump technique if $|a_f\rangle$
is selected to be a different hyperfine state \cite{zhu2011,Tewari}.
Furthermore, this rf spectroscopy technique works as an analog of
the powerful scanning tunneling microscope for probing the atomic
gases \cite{Shin,Jiang}: another atom will occupy the soliton state
after the original atom is scattered out by the probing laser.
Therefore, although there is only one atom in the kink and anti-kink
at a time, the population in $|a_f\rangle$ increases with the
probing time. Therefore, this scheme can be easily implemented in a
practical experiment. In this method, FPN can be indirectly deduced
from the population of the soliton state and then it is actually an
indirect method to measure FPN.


\section{discussion and conclusion}

Before concluding this paper, we discuss briefly how to
realize the FPN in 2D and 3D relativistic quantum field theories
\cite{Jackiw2}. It has been shown that a 2D Dirac Hamiltonian with a
vortex-like spatially inhomogenous mass term also supports a
zero-energy mode with one-half fermion number
\cite{Chamon,Seradjeh,Ruegg,Jackiw2}. The wanted 2D SO coupling
acting as the kinetic term in the Dirac Hamiltonian can be generated
in the previous laser-atom interaction configuration, such as
$c_x\sigma_xp_x-c_z\sigma_zp_z$ with
$c_z=\frac{1}{2}\eta\hbar\kappa\sin^2\alpha$. Other kinds of SO
coupling terms can also be generated via appropriate optical
dressing \cite{Dalibard}. Another crucial step is to simulate the
position-varying mass term with vortex profiles \cite{Jackiw2}.
Specifically in this cold atom system, one needs $\Omega_L(x,z)$
with profiles of the vortex type in contrast to the kink type for 1D
cases shown in Eq. (\ref{ModelHam1}). Fortunately, the needed laser
fields with vortex type defects can be created by using the
so-called optical vortex technique \cite{OV}, which moreover has
been implemented in cold atomic gases in experiments \cite{Wright}.

FPN can also be present in 3D Dirac systems, where the topologically
non-trivial background field should be replaced by a 3D profile of a
magnetic monopole \cite{Jackiw-Rebbi,Jackiw2}. For the 3D cases, we
need the SO coupling term such as
$\sigma_xp_x+\sigma_yp_y+\sigma_zp_z$, which can be synthesized by
the atom-light-interaction scheme proposed in Ref. \cite{3DSOC}. The
wanted mass term with monopole profiles may be generated by using
electromagnetic field superpositions like those were used to induce
3D Skyrmions in atomic gases \cite{Ruostekoski}.

In summary, we have proposed an experimental scheme to realize the
fractionalization of particle number with a 1D SO-coupled
ultracold Fermi gas. A kink-like potential and a
conjugation-symmetry breaking mass term are constructed by proper
laser-atom interactions, leading to an effective low-energy
relativistic Dirac Hamiltonian with a topologically nontrivial
background field. As a result, a localized soliton mode emerges
near the kink, having an FPN which is generally irrational and
experimentally tunable. The proposed cold atomic system provides a
direct quantum simulation of the famous generalized SSH model. We
have also presented two useful methods to detect the induced
soliton modes and the FPN in the system. In view of the fact that
SO-coupled Fermi gases were realized in two very recent
experiments \cite{Wang2012,Cheuk2012}, it is anticipated that the
present proposal will be tested experimentally in near future.

\section{acknowledgments}
We thank C.-Y. Hou, C. Chamon and Y. Chen for helpful discussions.
This work was supported by the NSFC (Grants No. 11125417, No.
10974059, No. 11104085, No. 91121023, and No. 11004065), the SKPBR
(Grant No. 2011CB922104), the GRF (HKU7058/11P) and CRF
(HKU-8/11G) of the RGC of Hong Kong.


\begin{thebibliography}{10}
\bibitem{note1} This Hamiltonian is corresponding to the Lagrangian $\mathcal{L}=i\overline{\Psi}\partial
\!\!\!/\Psi-\overline{\Psi}(\varphi_1+i\gamma^5\varphi_2)\Psi$ in
Ref. \cite{Goldstone}, where $\Psi$ is the field operator.
\bibitem{note2} The atomic transverse motion can be tightly confined by an
optical trap \cite{Liu}. Here we need the transverse trapping
frequency to be much larger than the SO interaction energy
$E_{\text{SO}}=mc_x^2/2$ with $c_x$ being given in Eq.
(\ref{JackiwHam}) and the energy gap $E_g$ being given in Sec. II.B.
Thus we can approximately consider a single transverse mode
throughout the text.
\bibitem{note3} There is an additional constant a.c.-Stark shift term $\left(
|\Omega _{y}|^{2}/\Delta_d \right) \left( |1\rangle \left\langle
1\right\vert +|3\rangle \left\langle 3\right\vert \right)$
associated with the laser beams $\Omega _{1y},\Omega _{3y}$. This
frequency shift is typically in the range from several kelohertz
to several hundred kelohertz, and is quite smaller compared to
frequency difference $\omega_{12}$ [cf. Fig. 2(a)], which is $228$
MHz for $^6$Li atoms and $1186$ MHz for $^{40}$K atoms. Thus it
has little physical effect on the ground levels and modifies
negligiblely the previous selection of laser beam $\Omega_L$ with
opposite detuning for states $|1\rangle$ and $|2\rangle$.
\end{thebibliography}
\end{document}